\documentclass{article}

\usepackage{tikz}
\usetikzlibrary{shapes,arrows,shadows,decorations,patterns}
\usetikzlibrary{
decorations.pathmorphing,calc,
backgrounds}
\usepackage{xcolor}
\usepackage{graphicx}
\usepackage{comment}
\usepackage[affil-it]{authblk}
\usepackage{cite}

\newcommand{\newpar}{\vspace{0.3cm} \noindent}
\newcommand{\indep}{\mbox{$\,\perp\!\!\!\perp\,$}}

\newcommand{\D}
{${\cal D}^{\tt EVD}_{\tt Outc}$}

\title{\huge \bf Causal Graph Aided Causal Discovery in an Observational Aneurysmal Subarachnoid Hemorrhage Study}

\author[a*]{\Large Carlo Berzuini}
\author[b]{\Large Davide Luciani}
\author[c]{\Large Hiren C. Patel}
\affil[a]{Centre of Biostatistics, School of Health Sciences, The University of Manchester, Oxford Road, Manchester, M13 9PL, UK}
\affil[b]{Unit\'a di Ingegneria della Conoscenza Clinica, Mario Negri Institute, Milan, Italy}
\affil[c]{Manchester Centre for Clinical Neurosciences, Stott Lane, Manchester, M6 8HD, UK\\

\vspace{0.2cm}
{\em *Corresponding author}: Carlo Berzuini (carlo.berzuini@manchester.ac.uk)}

\begin{document}

\maketitle

\begin{abstract}
\newpar Causal inference methods for observational data are increasingly recognized as a valuable complement to randomized clinical trials (RCTs).
They can, 
under strong assumptions, emulate RCTs or help refine their focus. Our approach to causal inference uses causal directed acyclic graphs (DAGs). 
We are motivated by a concern that many observational studies in
medicine begin without a clear definition of their objectives,
without awareness of 
scientific potential,
and without tools
to identify the necessary
in itinere adjustments. We 
present and illustrate
methods that provide "midway insights" during study's course, identify meaningful causal questions within the study's reach and point to the necessary data base enhancements for these questions to be meaningfully tackled. The method hinges on concepts of identification and positivity.  Concepts are illustrated through an analysis of data generated by patients with aneurysmal Subarachnoid Hemorrhage (aSAH) halfway
through a study, focusing in particular on the consequences of external ventricular drain (EVD) in strata of the aSAH population. In addition, we
propose a method for multicenter studies, to monitor the impact of changes in practice at an individual center's level, by leveraging principles of instrumental variable (IV) inference. {\bf Keywords:} causality; causal directed acyclic graph; DAG; randomized clinical trials; therapeutic decision
making; causal discovery; aneurysmal subarachnoid hemorrhage; instrumental variable; propensity score; conditional independence; observational studies;
external ventricular drain.
\end{abstract}

\fontsize{13pt}{15pt}\selectfont

\vspace{1.9cm}

\section*{\huge Introduction}

\vspace{0.2cm}

\newpar If every discourse originates from an assumption, a postulate we do not engage in proving, enclosed like an embryo in the yolk and the yolk in the egg, let the assumption at the root of the
present work be: \textsc{\small MEDICAL SCIENCE
IS BASED ON CAUSAL KNOWLEDGE}.

\newpar Randomized clinical trials (RCTs)—widely regarded as the "gold standard" for establishing causality—are increasingly facing challenges in addressing the complex causal questions that arise in therapeutic contexts. This has led to a growing interest in complementing RCTs with causal inference methods applied to observational data. These methods, under certain strong assumptions, can emulate the effects of RCTs or enhance their design and focus. Our approach to causal inference uses
causal directed acyclic graphs (DAGs)
\cite{greenland1999causal} \cite{dawid2024causal} to represent assumptions about the problem. In a causal DAG, nodes represent domain variables and arrows may represent causal relationships.
Under the interpretation adopted in this paper, the DAG simultaneously encodes causal and conditional independence assumptions about the variables. Thanks to this dual encoding, the causal graph enables us to efficiently utilize data to address causal questions that extend beyond the scope of purely predictive studies,
as a complement, or as an alternative, to RCTs.

\newpar Within this framework,
our method aims to provide "midway insights" during study's course, identifying clinically relevant causal questions within study's reach and the necessary data base enhancements for these questions to be meaningfully tackled. The method hinges on the
concepts of identification and positivity. The former ensures that the relationships depicted by the graph entail a unique, unbiased estimate of the unknown causal quantities,
irrespective of the model quantitative specification. Positivity ensures that the data contain sufficient information to allow for estimation of the causal parameters of interest. We emphasise
the method's applicability as data gradually become available throughout the study, for informed decisions to be made in selecting the study's appropriate direction.

\newpar We are concerned that too many observational studies in medicine begin without a clear definition of their objectives and with lack of awareness of the potential achievements. Hence the risk of wasting resources without a commensurate progress in causal knowledge. Our methodology helps
remedying this.

\newpar We illustrate
the conceptual points throught their application in an analysis of
data from 
patients with aneurysmal Subarachnoid Hemorrhage (aSAH), within a realistic midway study scenario. For illustrative
purposes, we have selected a scientific
question concerning the consequences
of a clinical intervention known as external ventricular drainage (EVD)
in specific strata of the population.

\newpar The next section provides an overview of therapeutic questions 
in aSAH that remain to be solved.
Section "Causal Graph of a Simple aSAH scenario" reviews basic notions of causal graphs with the aid of our SAH study. Next we deal with identification,
with positivity
and propensity scoring \cite{rosenbaum1983propensity}.
Section "Causal analysis of the effect
of EVD" does what it
promises. Finally, Section "Monitoring" 
focuses on multicenter studies
and expands on instrumental variable (IV) principles to develop a method for tracking the impact of practice changes at an individual center level, as
an alternative causality exploration paradigm.

\vspace{1.9cm}

\section*{\huge Background on SAH}

\vspace{0.2cm}

\newpar aSAH is a type of stroke
due to a ruptured aneurysm, causing
bleeding in the brain's subarachnoid space.
It affects roughly 9 in 100000 people
each year
\cite{Lawton2017}.
With its tendency to occur early in life, primarily affecting people aged 50 to 60, with mortality rates as high as 40-50\% and a quarter of survivors facing lifestyle limitations, aSAH results in a greater overall loss of productive life than other types of stroke.

\newpar A recent report from the Research Guidelines Committee of the National Institute for Health and Care Excellence (NICE)
\cite{NICE2021} highlights the
"glaring" absence
of an adequate evidence base for managing aSAH patients,
also criticising the lack of reliable outcome prediction models, which sometimes leads to specialists withholding treatment from patients who might have recovered well despite initially poor conditions. There
is clearly mileage for research to improve aSAH patient outcomes.
Well known difficulties of implementing
RCTs in this area call for use
of observational data.
Hence, in the first place,
we need tools for identifying which therapeutic questions can be effectively addressed using observational data and under what conditions.

\newpar This paper proposes methods for such an aim. Purely for
purposes of illustration, we shall devote special attention to
assessment of the consequences of
extraventricular drainage (EVD) 
in specific strata of the aSAH patient population. Our data base currently contains aSAH patient-level data from a multicenter hospital registry started in 2011, spanning 15 UK centers and Ireland.

\newpar Before tackling the
problem, it may be helpful to review the  scoring systems that
stroke professionals use to measure a patient's consciousness level by condensing multiple aspects of the patient's status into a single score. In our illustrative analysis, for sake of simplicity, we refrained from treating the different aspects as separate variables, although this would allow a more efficient use of the available information.

\newpar There is a significant perceived margin for improvement in aSAH outcomes, as well articulated by a recent report from the Research Guidelines Committee of the National Institute for Health and Care Excellence (NICE)
\cite{NICE2021}. The report highlights the "glaring absence of an adequate evidence base for the management of aSAH patients." It also criticizes the lack of a reliable model for predicting outcomes to guide triage decisions, which sometimes leads to specialists unintentionally withholding treatment from patients who, despite initially poor conditions, could have experienced a favorable recovery.

\newpar Randomized clinical trials in aSAH
represent a challenge, our
require careful pilot observational
studies. Progress thus depends on determining which questions in aSAH management can be effectively addressed using observational data and under what conditions. When these conditions are met, we need methods to analyze the data, leading to the development of evidence-based criteria for patient admission to treatment and selection for specific interventions based on individual circumstances.

\newpar This paper proposes methods to achieve this goal. For illustrative purposes, we 
denote special attention to assessing the consequences of extraventricular drainage (EVD) in specific strata of the aSAH patient population. Our database currently includes aSAH patient-level data from a multicenter hospital registry started in 2011, encompassing 15 centers in the UK and Ireland.

\newpar Before addressing the problem, it may be helpful to review the scoring systems that stroke professionals use to measure a patient's consciousness level by condensing multiple aspects of the patient's status into a single score. In our illustrative analysis, for simplicity, we have refrained from treating these different aspects as separate variables, although doing so would allow for a more efficient use of the information.

\begin{table}[h!]
\centering
\scalebox{0.75}{
    \begin{tabular}{|p{4.5cm}|p{10cm}|}
        \hline
        \textbf{GOS Category} & \textbf{Description} \\ \hline
        5 - Death & Death due to the initial brain damage. It may be categorized based on whether it occurred before or after regaining consciousness, distinguishing initial recovery. \\ \hline
        4 - Persistent Vegetative State & Unresponsive and speechless for weeks or months post-injury, with sleep-wake cycles emerging after 2-3 weeks. \\ \hline
        3 - Severe Disability (conscious but disabled) & Requires daily support due to physical and/or mental disabilities. \\ \hline
        2 - Moderate Disability (disabled but independent) & Independent in daily life, able to manage self-care and daily activities, though possibly with significant family disruption. \\ \hline
        1 - Good Recovery & Resumption of normal life with possible minor neurological or psychological deficits. Return to work could be influenced by socioeconomic factors and personal circumstances. \\ \hline
    \end{tabular}
}
\caption{\small Glasgow Outcome Scale (GOS) Categories}
\label{tab:gos_categories}
\end{table}

\newpar The Glasgow Coma Scale (GCS) is a composite of assessments across three response categories: eye (1 to 4 from no eye opening to opening eyes spontaneously), verbal (1 to 5 from no verbal response to oriented and conversational), and motor (1 to 6 from no motor response to obeying commands). A score at the upper limit of 15 denotes normal consciousness, whereas the lower threshold of 3 reflects a profound unconscious state.
Next we have the {\em World Federation of
Neurosurgical Surgeons} (WFNS)
score, used in the figure above. It is defined as an ordinal
5-level factor, where grade 1 corresponds to $GCS=15$;
grade 2 to $13 \le GCS \le 14$
with focal neurological deficits;
grade 3 to $13 \le GCS \le 14$
without focal neurological deficits;
grade 4 to $7 \le GCS \le 12$
and grade 5 to $3 \le GCS \le 6$.
In our analysis, we redefine {\tt WFNS}
as a 6-level ordered factor (occasionally treated as
a continuous variable for analytic convenience),
by adding level 6, which we define as corresponding
to an orginal WFNS value of 5 in the absence
of pupil reaction. GCS provides a basis for the definition
of the {\em Glasgow Outcome Scale} (GOS)
as given in the marginside.

\vspace{1.9cm}

\section*{\huge Causal graph for
a simple aSAH scenario}

\newpar We begin by introducing basic concepts of causal graphs, followed by a simple application to illustrate these ideas.

\vspace{0.9cm}

\subsection*{Theory}

\newpar A DAG defines a conditional independence structure for its variables, thereby determining a
family of probability distributions over  them. Conditional independence relationships can be read off the graph using several criteria, 
including the $d$-separation criterion,
which we are now going to summarize
\cite{Dawid1979,Pearl1988}.

\newpar $d$-separation
rests on the concept of {\em path} along a graph,
defined as consisting of
a consecutive sequence of edges, regardless
of their directions, as for example: ${\tt B} \rightarrow
{\tt D} \leftarrow {\tt A}$. 
A path may or may not
contain a pair of arrows that
collide head-to-head,
$\rightarrow \circ \leftarrow$.
If it does, the node between
the two arrows
is called a {\em collider}. 
We distinguish between
"blocked" and "unblocked"
paths on the graph in relation
to a particular query. A path
is {\em blocked} if it
contains at least
one non-collider
that is part of the conditioning
set in the query,
or at least
one collider that 
does not belong to
the conditioning set nor
has a descendant in that set.
 If, conditional
on a generic node
(or set of nodes)
${\cal C}$,
there are no {\em un}blocked paths
between two
generic (sets of) nodes, ${\cal A}$ and ${\cal B}$, then we say that
${\cal C}$ $d$-separates  ${\cal A}$ from
${\cal B}$, and conclude
that,
under all the probability
distributions
represented by the graph, ${\cal A}$ and
${\cal B}$ are conditionally independent
given ${\cal C}$.
If, instead, ${\cal C}$ does {\em not} $d$-separate ${\cal A}$ from
${\cal B}$, then there may exist, within the family of distributions
represented by the DAG, one particular
distribution
where ${\cal A}$ and
${\cal B}$ are {\em not}
conditionally independent
given ${\cal C}$, and we can 
(with slight abuse
of the terminology) claim
that ${\cal A}$ and
${\cal B}$
are {\em not} conditionally
independent given
${\cal C}$.

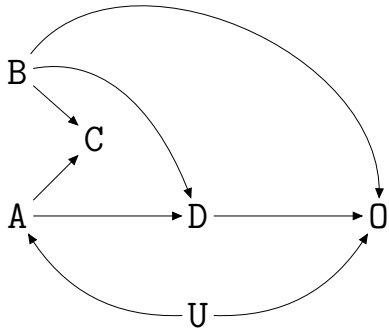
\begin{figure}[h]
\centering
\scalebox{0.6}{
\begin{tikzpicture}[align=center,node distance=4cm]
\node (e)[fill=none]{\Huge \tt D};
\node (e0)[left of=e,fill=none,node distance=4cm]{\Huge \tt A};
\node (c)[above of=e0,fill=none,node distance=3.2cm]{\Huge \tt B};
\node (a)[above right of=e0,fill=none,node distance=2.4cm]{\Huge \tt C};
\node (o)[right of=e,fill=none,node distance=4cm]{\Huge \tt O};
\node (u)[below of=e,fill=none,node distance=2.2cm]{\Huge \tt U};

\draw[-triangle 45] (e) to (o);
\draw[-triangle 45, bend left, out=75,in=110] (c) to (o);
\draw[-triangle 45] (e0) to (e);
\draw[-triangle 45, bend left] (u) to (e0);
\draw[-triangle 45, bend right] (u) to (o);
\draw[-triangle 45, bend left,out=50] (c) to (e);
\draw[-triangle 45] (c) to (a);
\draw[-triangle 45] (e0) to (a);
\end{tikzpicture}
}
\caption{\small Example of a directed acyclic graph (DAG).}
\label{fig:dag_example}
\end{figure}

\newpar For example
let's ask whether, according to the graph in Figure 1, variable {\tt B} is independent of {\tt A} once we are told the value of {\tt D}. In Dawid's notation, this assertion is written ${\tt B} \; \indep \; {\tt A} \; \mid \; {\tt D}$.
We find  via
$d$-separation that this assertion
does {\em not} follow from the graph,
meaning that conditional
on {\tt D}, variable {\tt B} 
is not independent of {\tt A}.
To justify this conclusion,
it suffices to consider that 
when we condition on {\tt D}  the
$ {\tt A}  \rightarrow {\tt D}
\leftarrow
 {\tt B}$ path
is unblocked.

\newpar As we stand at this point of the section, the DAG is seen as representing purely conditional independence properties
of a set of random variables, ie, a family of probability distributions over the graph.  Pearl
\cite{Pearl2009} would, however, imbue the DAG with causal significance in such a way that the same DAG will represent the causal properties of the system (the way the world works) and its conditional independence properties. In Pearl's formulation, this is possible by assuming
that the conditional distribution of any given node, given its parents, remains the same when we (hypothetically) switch from the observational setting where we collected the data to a setting where the values of some or all of the variables in the graph are changed by external intervention (interventional setting). Consider the following example. Interpreted causally, the DAG of Figure 1 tells us, for example, that the distribution of {\tt C}, once we know {\tt A} and {\tt B}, is unaffected by {\em how} the values {\tt A} and {\tt B} arose, whether in the same circumstances in which we collected the data, or by external intervention. To the extent to which these assumptions concern effects of interventions, they cannot be corroborated by observational data that are by definition collected free from interventions. As claimed by Dawid
\cite{Dawid2010}, these considerations impose limits on the machine-learning notion of "causal discovery", which attempts  to deduce causal properties solely from conditional independence relationships inferred from observational data. In contrast, our advocated approach makes "causal discoveries" by combining observational data information with pre-existing causal knowledge.

\vspace{0.9cm}

\subsection*{Application}

\newpar \textsc{\small Now pay
attention!} Let's illustrate DAG concepts using the straightforward -- perhaps overly simplistic -- scenario shown in Figure 2. Imagine an individual who has suffered from aSAH and, as a result, has been referred to a neurosurgical center for treatment. But which center? We know it. That’s indicated by the variable labeled {\tt Centre}. Once the patient arrives, various pieces of information are gathered from them, which we then represent in the graph as part of the node {\tt Pre-decision data}.

\newpar The gathered information is used by  doctors to decide whether the patient should
be admitted for treatment.
They may not be admitted, or left under
observation until perhaps their conditions improve enough to convince
the doctors they should be admitted.
This is how the {\em triage} goes about. And there's so much to say about triage -- but let's save that for the next paper! If the patient is admitted, the next step, depending also early in-hospital events, is to evaluate whether an EVD intervention is necessary.
This intervention
is represented in the graph
by node {\tt EVD}. At a subsequent landmark time or event, such as discharge or a fixed time post-admission, the patient's clinical outcome is observed. Arrows
are consistent with the direction of time, but wait
for a deeper definition 
of their meaning.
The following variables are involved:

\begin{figure}[h]
\centering
\scalebox{0.6}{
\begin{tikzpicture}[align=center,node distance=4cm]
\node (e)[fill=none]{\LARGE \tt EVD};
\node (e0)[left of=e,fill=none,node distance=4cm]{\LARGE \tt Pre-decision \\ \LARGE \tt data};
\node (c)[above of=e0,fill=none,node distance=3.2cm]{\LARGE \tt Centre};
\node (a)[above right of=e0,fill=none,node distance=2.4cm]{\LARGE \tt Admitted};
\node (o)[right of=e,fill=none,node distance=4cm]{\LARGE \tt Outcome};
\node (u)[below of=e,fill=none,node distance=2.5cm]{\Huge \tt U};

\draw[-triangle 45] (e) to (o);
\draw[-triangle 45, bend left, out=75,in=110] (c) to (o);
\draw[-triangle 45] (e0) to (e);
\draw[-triangle 45, bend left] (u) to (e0);
\draw[-triangle 45, bend right] (u) to (o);
\draw[-triangle 45, bend left, out=50] (c) to (e);
\draw[-triangle 45] (c) to (a);
\draw[-triangle 45] (e0) to (a);
\draw[-triangle 45, bend right] (e0) to (o);
\draw[-triangle 45, bend left] (a) to (o);
\draw[-triangle 45] (a) to (e);
\end{tikzpicture}
}
\caption{\small Causal DAG representation of a simple aSAH scenario.}
\label{fig:asah_dag}
\end{figure}
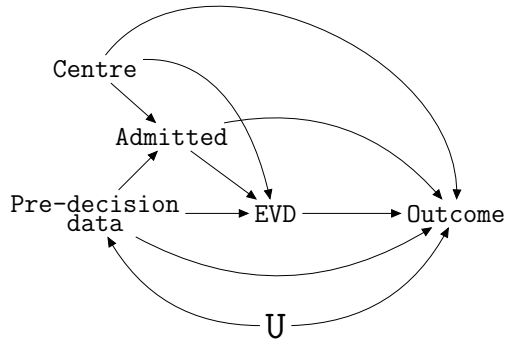

\begin{description}
    \item[{\tt EVD}:] A random binary variable, where 1 indicates that the doctors decide to perform the EVD intervention on the patient, and 0 indicates that the intervention is forgone.
    
    \item[{\tt Centre}:] A pointer to the surgical center where the patient has been referred for treatment.
    
    \item[{\tt Pre-decision data}:] 
everything relevant (excluding {\tt Centre}) in the patient's history 
up to EVD decision. A list of these
variables is given in Section "Causal analysis of the effect of EVD". Teo variables which will play an important role in the discussion are derived from the patient’s brain scan: {\tt Devastating scan}, which is self-explanatory, and {\tt AB}, defined as the ratio of the maximum width of the frontal horns of the lateral ventricles to the maximum internal diameter of the skull at the same level.
    
    \item[{\tt Outcome}:] A binary random variable where 1 indicates an unfavourable clinical outcome (as defined in the study) and 0 otherwise.
    
    \item[$U$:] The patient's unobserved, deep state of health.
    
    \item[{\tt Admitted}:] A binary indicator where {\em TRUE} indicates that the patient is admitted, and {\em FALSE} otherwise. Our
illustrative analysis conditions
on this node being {\em TRUE}
in all patients.
\end{description}

\newpar Some nodes represent
doctors' decisions. Others
characterise the patient's health status. Some could conceivably be intervened upon. For instance, {\tt Centre} decision is "intervenable" since it is possible to imagine intervening in the decision about the centre a patient is
referred to. Different centers may vary in their practice styles, not merely due to differences in patient populations but also because of professional preferences or differing opinions on the effects of interventions. Consequently, 
intervening on center decision may influence subsequent clinical
decisions and outcomes,
and this is represented in the graph
by causal arrows originating from {\tt Centre} and pointing towards {\tt EVD} and {\tt Outcome}.

\newpar Node {\tt Pre-decision data} contains
variables that may 
{\em inform} decisions
(admission
and or EVD) as well as
intervenable variables (eg {\tt Hypertension} that
may be causal to the outcome. If
we were to
expose the "guts"
of node {\tt Pre-decision data}, we would see details of its internal
causal relationships and of
its causal/informational
relationships with the remaining nodes of the graph. Our model also
allows some or all the pre-decision variables 
to directly correlate with
the patient's deep state of health, {\tt U}.

\newpar Note in our graph
that {\tt Pre-decision data} and {\tt Centre} are marginally independent but become dependent when conditioning on admission. This aligns with the observation that center-specific catchment areas exhibit similar frequencies of medical variables, while patient groups may differ with
respect to them due to differing admission criteria.

\vspace{2.9cm}

\section*{\huge Identification}

\vspace{0.9cm}

\subsection*{Theory}

\newpar Consider a causal DAG model ${\cal G}$ and let ${\cal V}$ denote the
set of variables of the dag that we have
observed and made available to the analyst. Let the causal question of interest ask whether  {\tt X} causes {\tt Y},
with ({\tt X, Y}) $\subseteq {\cal V}$. A crucial step in addressing the question is to determine whether, under the assumptions
encoded in ${\cal G}$, the information provided by
${\cal V}$ determines a unique, unbiased, estimate of the effect of {\tt X} on {\tt Y}. Pearl's {\em back-door} criterion gives a sufficient condition for identifiability. It requires the existence of a (possibly empty) subset ${\cal Z}$ of ${\cal V}$ such that no member of ${\cal Z}$ is a descendant of $X$ and
such that ${\cal Z}$ blocks 
(in the sense of $d$-separation) every path between $X$ and $Y$ that contains an arrow into $X$, so called
{\em back-door path}.

\begin{figure}
\vspace{0cm}
\begin{center}
\scalebox{0.65}{
\begin{tikzpicture}[align=center,node distance=4cm]
\node (e)[fill=none]{\LARGE \tt EVD};
\node (e0)[left of=e,fill=none,
node distance=4cm]{\LARGE \tt Smoking};
\node (c)[above of=e0,fill=none,
node distance=3.2cm]{\LARGE
\tt Centre};
\node (a)[above right of=e0,fill=none,node distance=2.4cm]{\LARGE \tt
Admitted};
\node (o)[right of=e,fill=none,
node distance= 4cm]
{\LARGE \tt Outcome};
fill=none,
node distance= 2.5cm]{\Huge \tt U};
\draw[-triangle 45] (e) to node [] {} (o);
\draw[-triangle 45, bend left, out=75,in=110] (c) to node [pos=0.65] {\color{gray}} (o);
\draw[-triangle 45, bend left,
out=50] (c) to node [pos=0.65] {\color{gray}} (e);
\draw[-triangle 45] (c) to node [pos=0.65] {\color{gray}} (a);
\draw[-triangle 45] (e0) to node [pos=0.65] {\color{gray}} (a);
\draw[-triangle 45, bend right] (e0) to node [pos=0.65] {\color{gray}} (o);
\draw[-triangle 45, bend left] (a) to node [pos=0.65] {\color{gray}} (o);
\draw[-triangle 45] (a) to node [pos=0.65] {\color{gray}} (e);
\end{tikzpicture}
}
\caption{\small The 
causal effect of antecedent smoking
on outcome is 
(nonparametrically) identified
by the set of variables represented
in the graph, with {\tt Admitted} $=$
TRUE, based on a sample of
admitted patients.\\}
\end{center}
\end{figure}
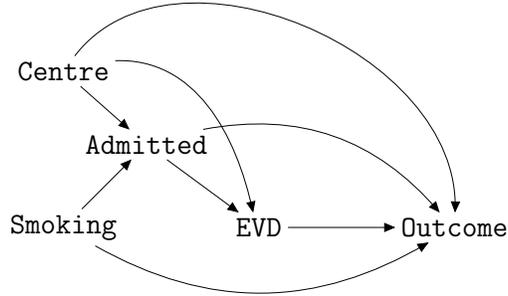

\vspace{0.9cm}

\subsection*{Application}

\newpar This section
demonstrates the capability of the
DAG machinery, in conjunction with $d$-separation, to determine whether a causal question is nonparametrically identified in a given
analysis situation. We do it through
examples.

\newpar Our first example 
considers the question
whether the variables
represented in the graph of Figure 3
identify the causal effect of 
antecedent {\tt Smoking} habit
on {\tt Outcome}, in an admitted patient,
under the assumptions represented in the graph. The answer to the question
is yes. In fact, once
we condition on {\tt Admitted}$= TRUE$ and {\tt Centre}, no unblocked
back-door paths
exist between {\tt Smoking} and {\tt
Outcome}.

\newpar Figure 4 represents an
elaboration of the graph
of Figure 3, involving the addition
of node {\tt Hypertension}. The 
causal effect of antecedent smoking
on outcome is no longer
(nonparametrically) identified,
whether or not we condition
on {\tt Hypertension}. In fact,
in this new scenario, we
have two potential back-door
paths with respect to the effect of
{\tt Smoking}: The
first is {\tt Smoking}
$\rightarrow$ {\tt Admitted $=$ TRUE}
$\leftarrow$ {\tt Hypertension}
$\leftarrow$ {\tt U} $\rightarrow$
{\tt Outcome}.
The second is
{\tt Smoking}
$\rightarrow$ {\tt Hypertension}
$\leftarrow$ {\tt U} $\rightarrow$
{\tt Outcome}. The former is
unblocked when we do {\em not}
condition on {\tt Hypertension}.
The latter is unblocked when
we {\em do}
condition. Luckily, the
problem may be successfully tackled via
inverse probability weighting
\cite{robins2000marginal}.

\newpar We've just seen
that adding the {\tt Hypertension} node to the graph in Figure 3 disrupts the identifiability of the effect of prior smoking. This underscores the fragility of our assumptions when dealing
with our uncertainty about the biological universe is challenging.

\newpar Let us now focus
on the causal effect of
{\tt EVD} on {\tt Outcome}. Let
the symbol
 \D 
 denote the set of variables in
the graph such that conditioning
on (\D, {\tt Admitted}$=$ TRUE,
{\tt Centre})
blocks all back-door paths
between {\tt EVD} and {\tt Outcome}.
Variables in \D are called \D-variables.
When all \D-variables are available
to the analyst we have \D-completeness.
In this case
we can estimate the causal effect
of {\tt EVD} on {\tt Outcome} based
on a sample of admitted aSAH patients,
adjusting for \D and, where appropriate, for {\tt Centre}. Methods of matching based on propensity
scoring are recommended.

\newpar With reference
to Figure 4, we have
\D = ({\tt Smoking},
{\tt Hypertension}).
Under \D-completeness, ie, when these
two variables are available,
a regression
analysis of the dependence
of {\tt Outcome} on
{\tt EVD}, within a sample
of admitted patients,
provided the variables  {\tt Centre},
{\tt Smoking} and {\tt Hypertension}
are adjusted for,
will estimate
the causal effect of
{\tt EVD} on {\tt Outcome}.

\begin{figure}
\begin{center}
\scalebox{0.4}{
\begin{tikzpicture}[align=center,node distance=4cm]
\node (e)[fill=none]{\LARGE \tt EVD};
\node (e0)[left of=e,fill=none,
node distance=8cm]{\LARGE \tt Smoking};
\node (hyp)[left of=e,fill=none,
node distance=4cm]{\LARGE \tt Hypertension};
\node (c)[above of=e0,fill=none,
node distance=3.2cm]{\LARGE
\tt Centre};
\node (a)[above right of=e0,fill=none,node distance=2.4cm]{\LARGE \tt
Admitted};
\node (o)[right of=e,fill=none,
node distance= 4cm]
{\LARGE \tt Outcome};
\node (u)[below of =e,
fill=none,
node distance= 2.5cm]{\Huge \tt U};
\draw[-triangle 45] (e) to node [] {} (o);
\draw[-triangle 45, bend left, out=75,in=110] (c) to node [pos=0.65] {\color{gray}} (o);
\draw[-triangle 45, bend left] (u) to node [pos=0.65] {\color{gray}} (hyp);
\draw[-triangle 45, bend right] (u) to node [pos=0.65] {\color{gray}} (o);
\draw[-triangle 45, bend left,
out=50] (c) to node [pos=0.65] {\color{gray}} (e);
\draw[-triangle 45] (c) to node [pos=0.65] {\color{gray}} (a);
\draw[-triangle 45] (e0) to node [pos=0.65] {\color{gray}} (a);
\draw[-triangle 45, bend right,
out=-90, in=-90] (e0) to node [pos=0.65] {\color{gray}} (o);
\draw[-triangle 45, bend left] (a) to node [pos=0.65] {\color{gray}} (o);
\draw[-triangle 45] (a) to node [pos=0.65] {\color{gray}} (e);
\draw[-triangle 45] (hyp) to node [pos=0.65] {\color{gray}} (e);
\draw[-triangle 45] (e0) to node [pos=0.65] {\color{gray}} (hyp);
\draw[-triangle 45] (hyp) to node [pos=0.65] {\color{gray}} (a);
\end{tikzpicture}
}
\caption{\small 
Slightly more elaborate version of the preceding figure. The 
causal effect of antecedent smoking
on outcome is no longer
(nonparametrically) identified,
whether or not we condition
on {\tt Hypertension}.}
\end{center}
\end{figure}
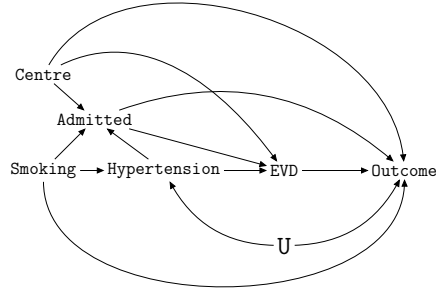

\vspace{1.9cm}

\section*{\huge Positivity}

\vspace{0.3cm}

\newpar Let's assume the causal
effect of interest is identified
by the data we have. This doesn't necessarily guarantee the data
{\em contains information} about the effect. We are
now going to address this issue
by introducing and then using the concept of {\em positivity}.

\newpar Let the symbol $a$ denote the generic value of
the intervention variable $A$ with domain ${\cal A}= (0,1)$ say, where $0$ indicates
"no intervention" and $1$ "intervention".
Let
the symbol
$pa_A$ denote the set of variables
that are parents of $A$ in the graph,
including {\tt Centre} and all those
variables in
{\tt Pre-decision data} that
inform the EVD decision.
Let the symbol
${\cal S}_A$ denote the region
of values of $pa_A$ in the patient
(sub-)population of interest.
Let the symbol ${\cal D} $ denote
the data available to the
analyst. Then a
necessary condition for the effect of the intervention on patients
in ${\cal S}_A$ to be
estimable from the data is
\cite{dawid2010identifying}:

\begin{quote}
\newpar {\bf Positivity condition for 
evaluating the effect of
intervention $A$ on patients
in  ${\cal S}_A$):
for every configuration
$pa_A = pa_A^{*}$, 
with $pa_A^{*} \in {\cal S}_A$, 
and for each $a \in {\cal A}$, we have
\begin{equation}
\label{positivity}
p(a \; \mid \; pa_A^{*}; {\cal D}) > 0,
\end{equation}
\noindent where the probabilities
involved are
large enough
for the data to provide a sufficient amount of information to estimate the
effect of $A$.}
\end{quote}

\newpar The above condition
tells us that for every possible
configuration of the variables that inform
the decision about the intervention and
of the intervention variable, there
is a positive probability of finding
a patient with that configuration in the
subpopulation of interest.
This implies
that if we regard
the probability
$b(pa_A) \equiv p(a=1 \; \mid \; pa_A; {\cal D})$ as a scalar
function of $pa_A$, called the {\em balancing score},
and we consider, for each intervention group,
its probability density
profile over this function, the profiles are
continuous with respect to each other,
meaning, in practice, that each profile rises
above zero wherever the other one does so.

\newpar The balancing score,
$b(pa_A)$, must
be estimated from the data. One possibility is to
assign the $b(.)$ function an {\em a priori} parametric
form with unknown parameters to be estimated from the
data. Since estimation will not involve the outcome, no
multiple comparisons bias will affect this procedure.
Machine learning algorithms to estimate $b(.)$ are welcome.

\vspace{2.9cm}

\section*{\huge Causal analysis of the effect of EVD}

\newpar The themes we've discussed so far will converge in this section, much like the spokes of an upside-down fan.

\newpar The analysis we are about
to present illustrates what
can be done midway through 
an observational study with the
aid of a causal graph, as an aid
to an effective management.

\newpar Data collection began in 2011, and the database now encompasses over 18000 aSAH patients, with near-complete recordings of 133 variables. These include: Centre, Age, Sex, Hypertension, Ischaemic Heart Disease, Smoking, Diabetes, Recreational Drug Use, Ictus Date, Referral Date, WFNS Grade on Referral, Fisher Grade, Admission Date, GCS on Admission to Neurosurgical Unit, Pupillary Response on Admission, Pre-admission Modified Rankin Score, Cardiovascular Stability, Ruptured Aneurysm Location, Date of Discharge, Discharge Destination, Length of Stay, Glasgow Outcome Score at Discharge, and Glasgow Outcome Scale at 3 months post-admission. Further variables, omitted from the above list, concern clinical decisions and interventions other than (subsequent to) EVD, such as aneurysm treatment. The aSAH therapy problem does, in fact, extend far beyond the confines of our graph models.

\newpar One author (HP) dedicated considerable effort to interviewing key "actors in the scene", specifically, the aSAH professionals responsible for daily triage and EVD decisions. These interviews resulted in a list of variables that stay "on top of the doctors' heads" during their
routine.  This list was then compared against the existing collection of database variables. A main discrepancy was discovered: the absence of brain scan (BS) information from the database,
and, in particular, of the previously
mentioned variables {\tt Devastating scan} and {\tt AB} ratio.

\newpar The perceived importance of 
the missing BS variables, affecting both EVD decision-making and the patient’s subsequent evolution, generated trepidation. BS information was clearly a potential unobserved confounder in our analysis. We faced two options: either start systematically digitizing and recording BS information on a long series of future patients or, if feasible, retrieve and digitize the brain scans of past patients and add the information to the database. The latter option proved viable. So far, we have added BS data for 258 patients, alongside the other listed variables.

\begin{figure}[h]
\begin{center}
\scalebox{0.7}{
\includegraphics[width=12.5cm,
height=10cm]{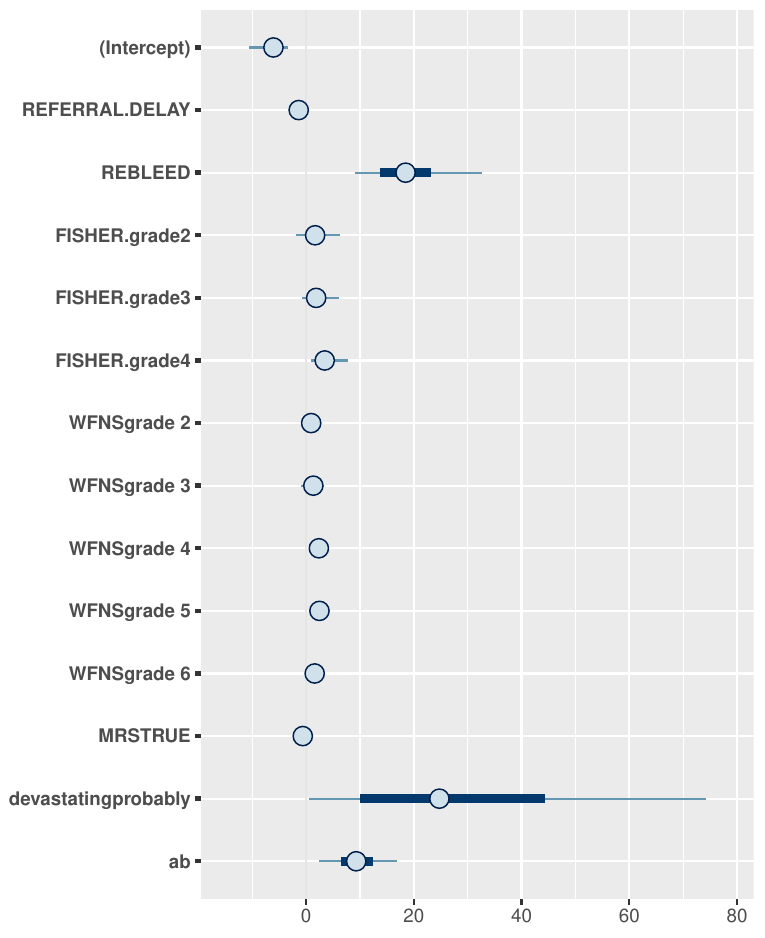}
}
\caption{\small Forest plot summary
of the fitting of a Bayesian logit-linear
regression model of the indicator
of sustained EVD on all its known
influences, based on the sample
of 258 patients complete
with brain scan information.
}
\end{center}
\end{figure}

\newpar Among BS variables, {\tt Devastating Scan} is self-explanatory,
and {\tt AB} represents the ratio of the maximum width of the frontal horns of the lateral ventricles to the maximum internal diameter of the skull at the same level. 

\newpar In the following description
of our pilot analysis of the 258 patients,
the goal is not to provide an exhaustive report of the results but rather to highlight the "causal discovery" idea.

\newpar A Bayesian logit-linear regression model of the binary intervention variable, {\tt EVD}, was fitted to the sample
of 258 patients considering all the
candidate influencing factors (including BS variables). Results from the fitting are summarized in the forest plot shown in Figure 5. In the plot, circles represent point estimates of covariate effects on the log-odds of 
sustaining EVD, while the extending lines indicate the corresponding 95\% Bayesian credible intervals. The plot demonstrates the dominant influence of brain scan variables and {\tt Rebleed} on the decision to administer EVD, with the contributions of the remaining variables appearing relatively minor in comparison.

\newpar We used the model-predicted probability of EVD, or {\em EVD propensity
score}, as a balancing score, ($b(.)$ in our
previous notation) between the two intervention groups (EVD vs. no-EVD). A stochastic matching algorithm
\cite{rosenbaum2007} was then applied to the original sample of patients to extract two smaller, well-matched intervention groups with respect
to this balancing score. Figure 6 illustrates the density profiles of the two intervention groups along the balancing score, both before (left plot) and after (right plot) matching. 

\newpar In both cases, the densities of the two groups do not have identical support, indicating a lack of positivity. 
The figure suggests that
within the aSAH patient population, there are "extreme ends" (very low or very high propensity) where clinicians generally agree on whether or not to perform EVD, but lack of consensus on EVD in the central region of the propensity score profiles. The lack of positivity signals potential challenges in assessing the overall effect of EVD in the aSAH population, even with comprehensive data collection, including BS information, from a large patient cohort.

\begin{figure}[h]
\begin{center}
\scalebox{0.7}{
\includegraphics[width=12.5cm,
height=10cm]{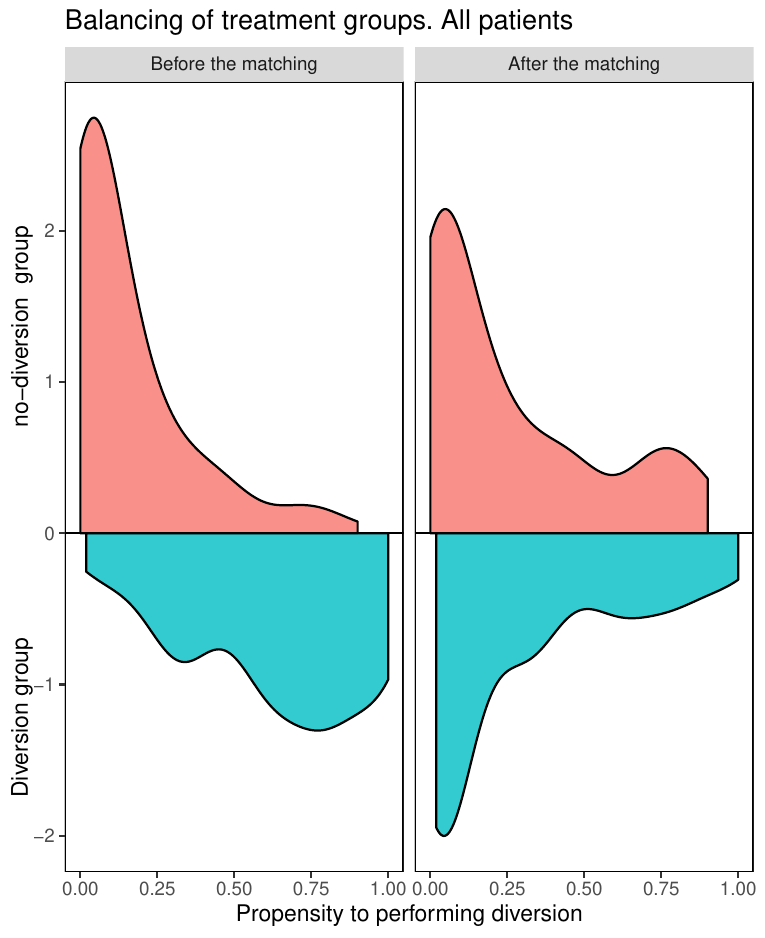}
}
\vspace{0.2cm}
\caption{small Assessment of balance between intervention 
(EVD vs no-EVD) groups 
with respect to propensity to EVD,
from an analysis of the 258 patients with available brain scan information.}
\end{center}
\end{figure}

\newpar Next, we focused on the 158 patients with WFNS grade 1 within the sample of 258 patients, trying
to glean from this limited sample an answer to the
question whether a comparison
of EVD versus no-EVD within this specific patient category is within
reach in our observational study. Figure 7 visualizes the balancing of the two intervention groups by juxtaposing their density profiles along the balancing score, calculated specifically for this category. The degree of balancing appears to be poor due to lack of positivity. Most patients in this stratum appear to have a very low propensity to receive EVD, and within this 
low-propensity group, there appears to be significant disagreement among doctors as to whether EVD should be performed. However, this category also includes a subgroup with extreme propensity for EVD, corresponding to a small density peak at the upper extreme of the horizontal axis. This is likely to correspond to grade 1 patients with poor brain scan results. What we learn from this plot is that evaluating the effect of EVD within a patient stratum not defined by brain scan variables is of limited value. In conclusion, the question of the effect of EVD in grade 1 patients is poorly framed.

\begin{figure}[h]
\begin{center}
\scalebox{0.7}{
\includegraphics[width=12.5cm,
height=10cm]{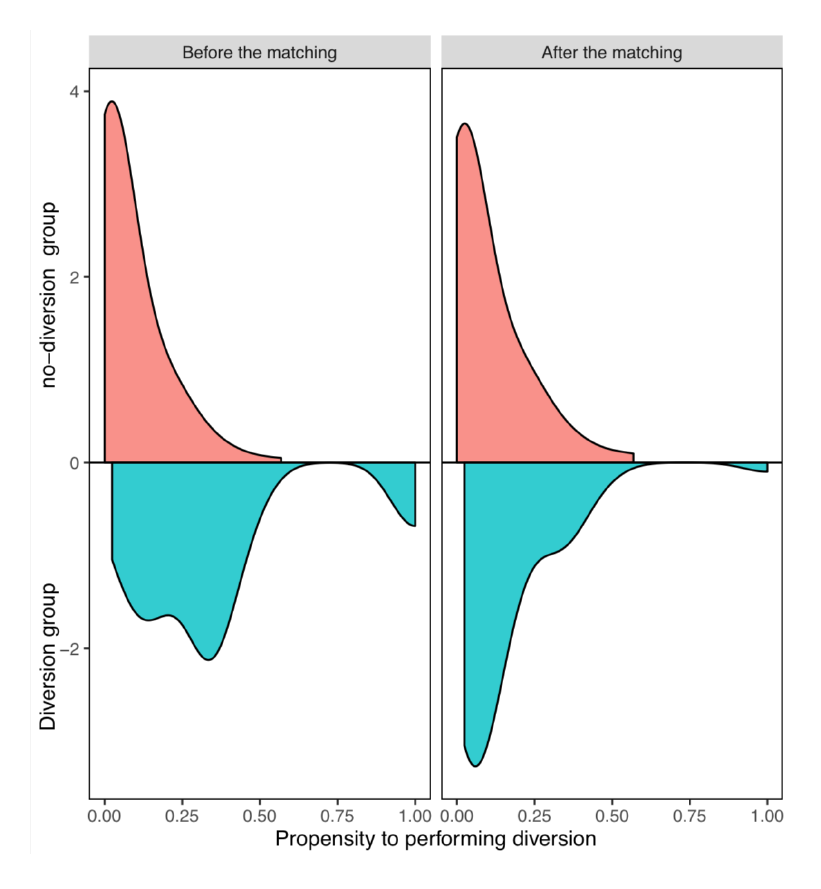}
}
\vspace{0.2cm}
\caption{\small Assessment of balance between EVD intervention groups 
with respect to propensity to EVD,
from an analysis of grade 1 patients with available brain scan information.
}
\end{center}
\end{figure}

\newpar We then moved to
considering the question:
\begin{quote}
{\em "is EVD
beneficial within
grade 1 patients
with no rebleeding
and an {\tt AB} ratio greater than
0.12?"
}
\end{quote}

\newpar Out of the 258
patients of our sample, 147
belonged to the category in question. 
Our analysis of the 147 patients produced
the balancing plot of Figure 8.
The degree of balancing
is good, thanks
to the uniform
disagreement among doctors/centres
about the best EVD choice 
for this
patient category. The post-matching
EVD vs no-EVD balance is, in fact, very good
in this category. 
In conclusion, it may be 
a good idea to pursue the above
question (or even better questions
posited by the specialists) once
the larger database has been completed
with brain scan data. 
Post-matching we remained
with a sample of 68 patients, 
corresponding to
a sampling ratio of 0.46.
As uncertain as this ratio is,
given the exiguity of the sample,
it nevertheless provides
a ballpark estimate of the sample size
required to emulate the power
of a randomized clinical trial of, say, 100
recruits. We would approximately need
100/0.46 =   218 grade 1 patients 
with no rebleeding and {\tt AB} ratio
greater than 0.12.

\begin{figure}[h]
\begin{center}
\scalebox{0.7}{
\includegraphics[width=12.5cm,
height=10cm]{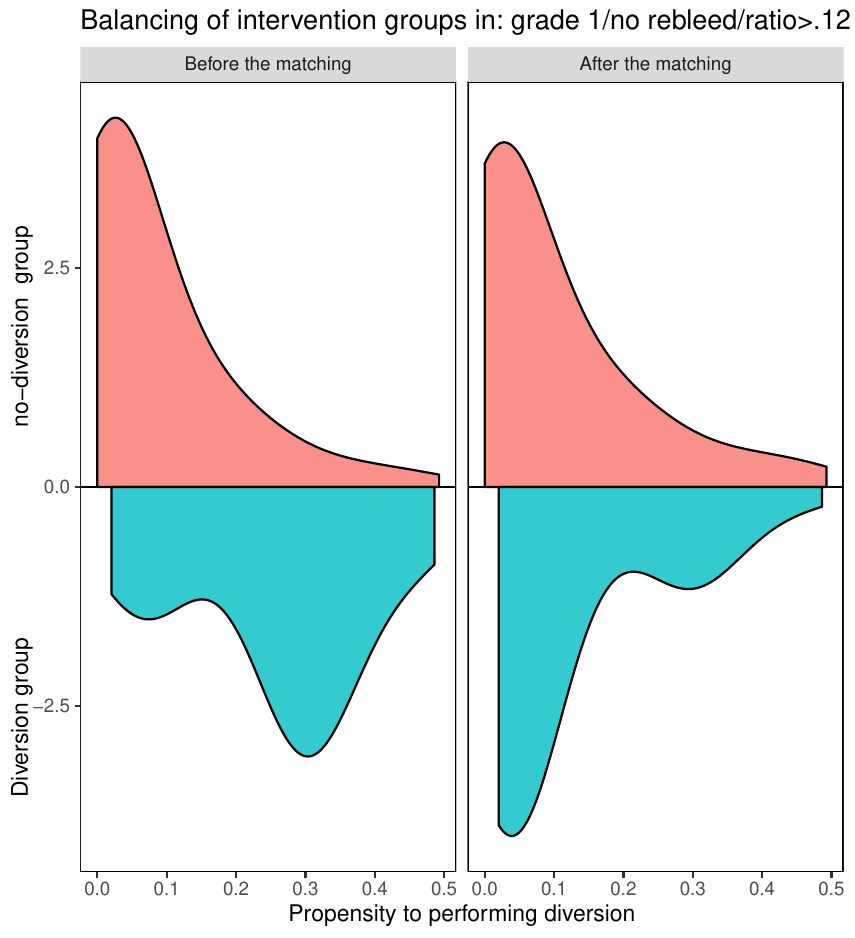}
}
\vspace{0.2cm}
\caption{\small Assessment of balance between EVD intervention groups 
with respect to propensity to EVD,
from an analysis of patients with available brain scan information,
characterized by a WFNS grade 1, no rebleeding
and an {\tt AB} ratio greater than
0.12.}
\end{center}
\end{figure}

\newpar The balancing plot
from an analysis of the patients
with $AB$
ratio greater than 0.5 is shown in Figure 9.
The balancing is poor, one reason being that 
the EVD decision is undisputed
at the extremes 
of the considered 
interval of $AB$ values, leaving only
some uncertainty and disagreement
in the central region of the 
propensity axis.

\newpar As a final example,
note in Figure 10 the good balancing of
the EVD intervention groups within
the group of patients
with $AB$ ratio between  0.1 and 0.5.
This plot
reveals lack of consensus about
the best EVD decision over the entire range
of EVD propensity. This is one question we might
wosh to pursue with a complete database.

\begin{figure}[h]
\begin{center}
\scalebox{0.7}{
\includegraphics[width=12.5cm,
height=10cm]{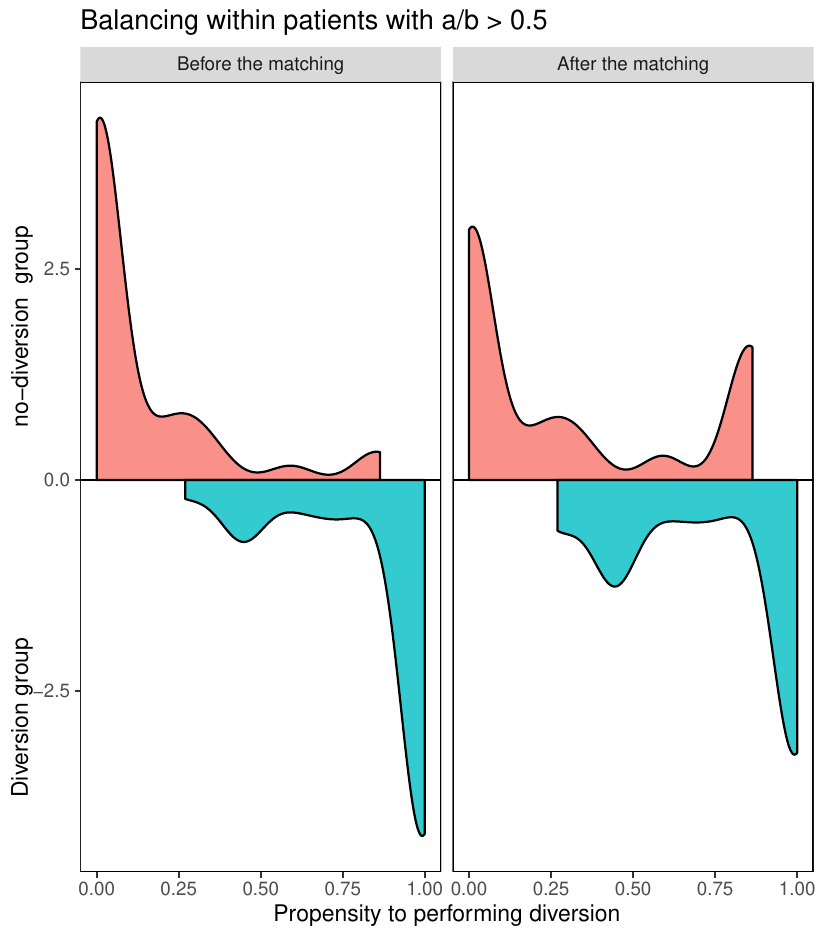}
}
\caption{\small Again, poor balancing of
the EVD intervention groups with
$AB > 0.5$.
}
\vspace{0.2cm}
\end{center}
\end{figure}

\begin{figure}[h]
\begin{center}
\scalebox{0.7}{
\includegraphics[width=12.5cm,
height=10cm]{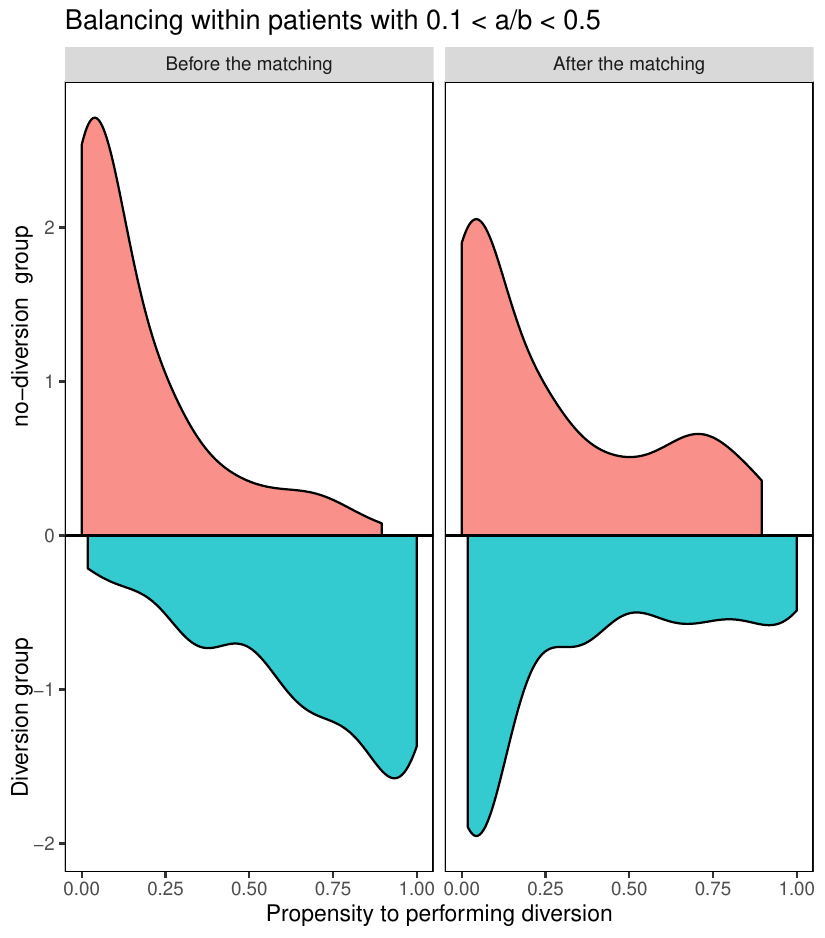}
}
\vspace{0.2cm}
\caption{Good Balancing of
the EVD intervention groups within
patients with
$AB$ ratio between  0.1 and 0.5.
}
\end{center}
\end{figure}

\newpar Some might find the questions raised in this section less compelling. 
Never mind! We hope we have illustrated
the ability of the method to highlight
the importance of adding
certain variables to the database, and to identify from a large number of potential causal questions, those that are actually feasible to pursue using observational data, once those variables have been incorporated.

\vspace{1.9cm}

\section*{\huge Monitoring}

\vspace{0.7cm}

\newpar In this final section, we
show results from an analysis of the
global sample of 18000 patients
with WFNS grade 3-5, 
across 18 participating centres, by omitting
brain scan variables due to their
huge percentage of missing values.
Figure 11 summarizes the results
from a  linear
regression analysis of the EVD indicator,
allowing a near-complete collection
of EVD influences (bar BS variables)
to enter the model. Variable {\tt Centre}, in particular,
contributed 17 unknown parameters
into the model, with Centre 1 taken
as reference.

\newpar Noteworthy in this figure is the remarkable 
amount of variation in propensity to EVD introduced by the variable
{\tt Centre}, compared with that of the most important variables characterizing the patient's health status. It is unlikely that this variation is an artifact due to missing brain scan information. A similar observation applies to Figure 12, which summarizes results from fitting a linear model of the probability of an adverse outcome using the same set of explanatory variables as the preceding regression. Joint inspection of Figures 11 and 12  suggests  proportionality between the effects exerted by each centre on EVD and on outcome. This
is important, because evidence that the centre effect on EVD propensity leads to a proportional effects on outcome
may, under certain assumptions, be  a sign of causality.

\newpar For a closer look into proportionality
we have generated the plot in Figure 13, where each $i$th
anonymized centre is 
represented
by a circle with  coordinates $(\alpha_i, \beta_i)$,
where $\alpha_i$ denotes the estimated centre effect on EVD propensity,
and $\beta_i$ the estimated centre effect on
probability of n unfavourable outcome.
The horizontal axis in Figure 12 quantifies the centre's 
negative propensity (reluctance) 
to administer EVD, and the vertical axis is the estimated centre-specific
probability of adverse outcomes.
Centres falling near the upper right corner of the plot are characterised
by low propensity to EVD and
correspondingly high risk of an adverse outcome.

\newpar Figure 13 is, indeed, mildly suggestive that performing EVD is beneficial for a poor grade patient. In fact, it
suggests a tendency of an unfavourable outcome to occur in
centres who are "reluctant" to perform EVD. This 
interpretation is emphasized by the positive
 slope of the superimposed line. which is, in fact, Egger's instrumental variable (IV) estimate of
the causal effect of avoiding EVD on probability of an unfavourable outcome,
and has been calculated by taking precision of estimates into account.
In other words, we are leveraging the diversity
of preference for EVD across centres as an instrument for
probing a possible causal effect of EVD on outcome.
Validity of this approach depends on untestable assumptions, including the
“instrument strength independent of direct effect” (InSIDE) assumption
that centre effects on outcome that are not mediated by EVD be uncorrelated with the corresponding effects on propensity to EVD. 

\newpar Due to this result
having been obtained  from data devoid
of brain scan patient information, we remain hesitant to offer
this as evidence in favour of a causal hypothesis. We
are just interested in illustrating the method. Note that
this plot offers more than just a numerical
effect estimate: it marks the position of each individual
centre with respect to the remaing ones in terms of outcome and
propensity to EVD, as a basis for 
centre performance monitoring in relation to 
the impact of specific aspects of the centres' clinical practices.

\newpar The two
regressions were fitted by using a  linear link function
(a choice made possible by the
limited range of values of
the probabilities of an unfavourable
outcome) rather than
a logistic one
to avoid problems of collapsibility.

\begin{figure}[h]
\begin{center}
\scalebox{0.7}{
\includegraphics[width=12.5cm,
height=14cm]{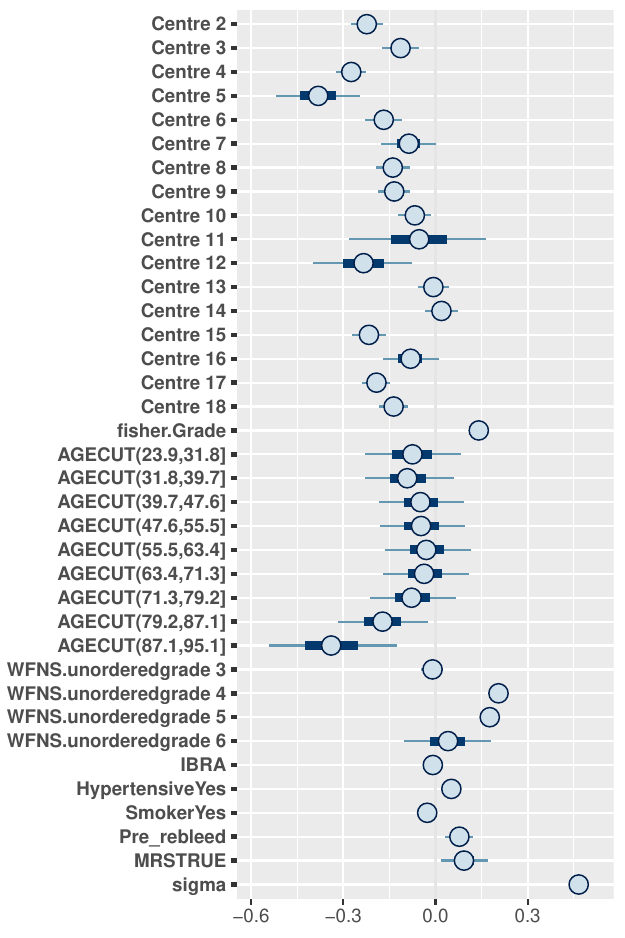}
}
\vspace{0.2cm}
\caption{\small Forest plot summary
of the fitting of a Bayesian linear
regression model of the indicator
of sustained EVD on 
its influences, including the 18-level factor {\tt Centre}.
}
\end{center}
\end{figure}

\begin{figure}[h]
\begin{center}
\scalebox{0.7}{
\includegraphics[width=12.5cm,
height=14cm]{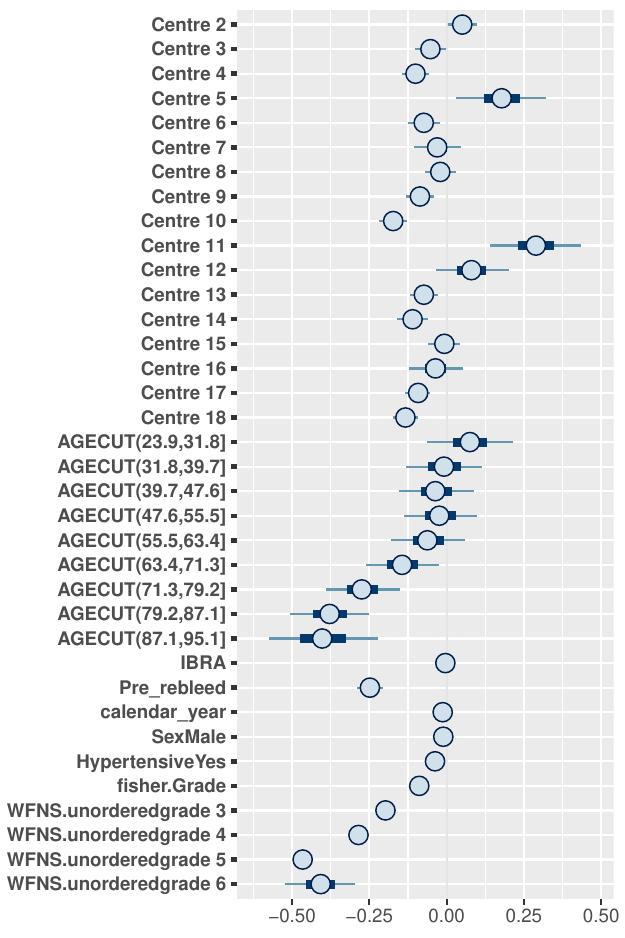}
}
\vspace{-0.2cm}
\caption{Forest plot summary
of the fitting of a Bayesian linear
regression model of the binary outcome variable on all known pre-decision patient variables, including the 18-level factor {\tt Centre}.
}
\end{center}
\end{figure}

\begin{figure}[h]
\begin{center}
\includegraphics[scale=0.25]{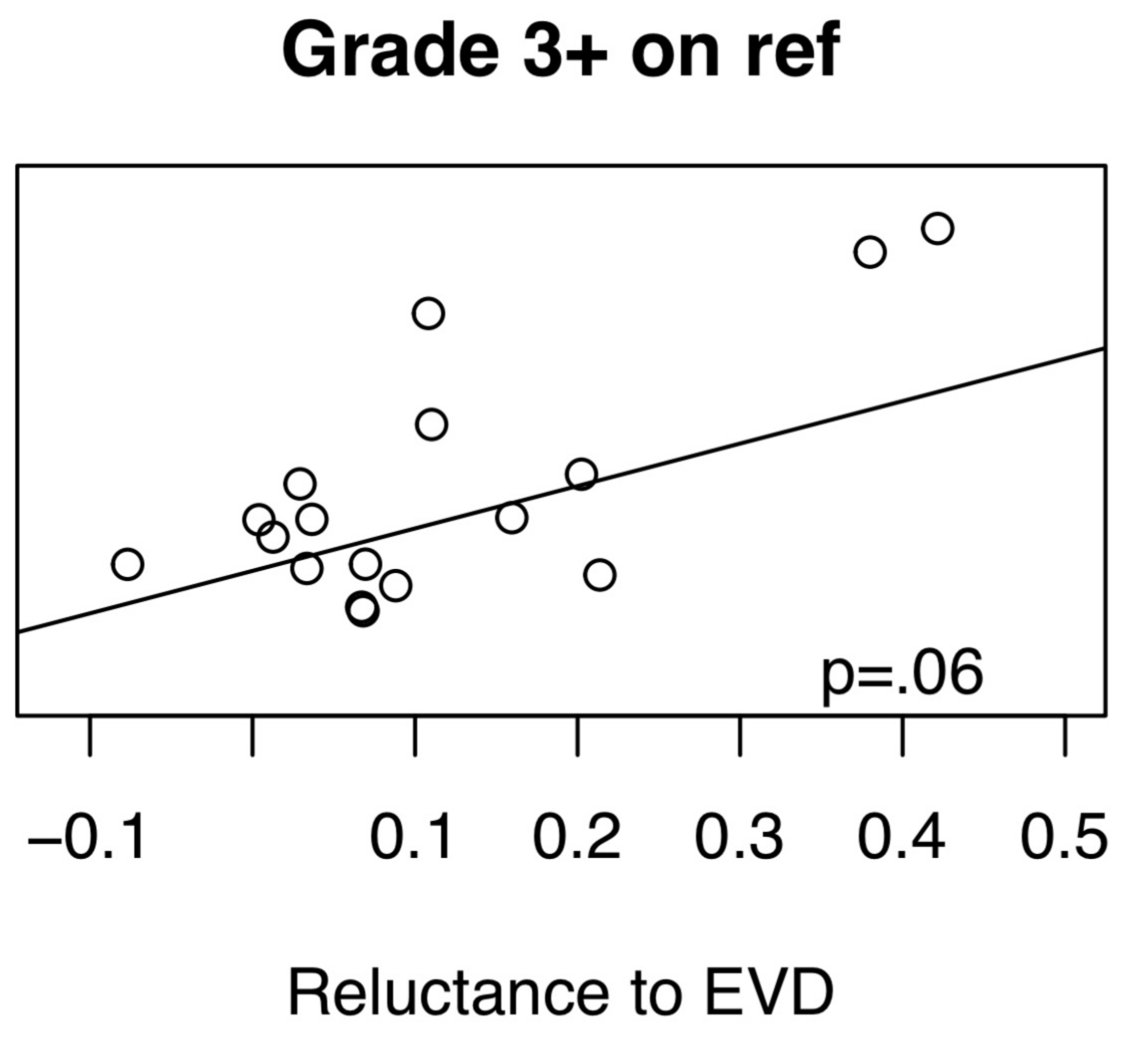}
\vspace{-0.2cm}
\caption{
\scriptsize Results from 
an analysis of patients with poor WFNS grade. Participating centers are 
represented by circles, with horizontal coordinate corresponding to the centre's 
negative propensity (reluctance) 
to perform EVD, and the vertical one representing the estimated centre-specific
probability of adverse outcomes.
The slope of the superimposed line equals Egger's IV estimate 
of the effect of
administering EVD in a poor
grade patient, calculated by taking precision of estimates into account.}
\end{center}
\end{figure}

\vspace{1.9cm}

\section*{\huge Discussion}

\newpar Unlike conventional machine-learning approaches, our "causal discovery" method makes explicit the necessary assumptions for drawing valid causal conclusions, empowering users to actively manage these assumptions through the use of a causal graph.

\newpar What we refer to as "causal discovery" in this paper is an evolving process. As the study progresses, incoming data are integrated with the causal graph to sharpen the focus on causal questions that can be effectively addressed using observational data. This approach also helps to identify the key areas where the database needs to be expanded to achieve the study's objectives.

\newpar Many clinical observational studies rely on high-quality hospital registries, which initially include only a limited set of clinical variables due to the high cost and time-consuming nature of data collection. As the study advances and specific scientific questions emerge, it becomes crucial to ensure that the registry contains all the necessary data to address these inquiries. Our method
involves conducting "in itinere" pilot studies (such as our study with 258 patients) and using the causal graph to identifying data gaps early on and providing targeted recommendations for selectively enriching the database.

\newpar In Section "Monitoring," we introduce a different approach to the problem. Rather than aiming for a quantitative assessment of causal effects, this approach visualizes the trajectories of individual treatment centers in terms of their propensity for specific interventions and their outcomes. The relative positions of centers on the plot offer insights into different styles of practice and their impacts on outcomes.

\newpar The case of aSAH serves as a valuable benchmark for applying the proposed methods in a real clinical setting. It is challenging to grasp the potential impact of our method without considering the sociological aspects of how the medical community advances research. aSAH is a fitting example because therapeutic decisions in aSAH are complex, with a lack of an adequate evidence base to guide practice, allowing considerable clinical freedom. In one scenario, clinicians might converge on a particular intervention policy unsupported by evidence, leading to a lack of positivity and research inertia. In another scenario, the medical community may split into different camps, that adopts intervention policies
at odds with each other. This also risks a lack of positivity, thereby hampering knowledge advancement through observational studies. This division can also prevent an RCT (randomized controlled trial) from being accepted due to difficulty in reaching consensus on equipoise. Our method allows researchers to use observational data to capture snapshots of the situation and use them to steer the study in the most promising direction.

\newpar One final remark concerns causal questions that are both intriguing and challenging to pursue. One such question is, "Who is expected to benefit from being admitted to treatment?" The question is important in
the light of recent criticisms of the current
aSAH triage regime. Such a question must be approached using data from individuals referred to various centers for aSAH treatment, where some have been admitted and others have not, with both groups followed up for an appropriate period before measuring outcomes. Randomizing admission is clearly not an option, so we must ask whether observational data can help. The answer is "yes ... but." If all centers use the same admission criteria, there will be insufficient positivity and the data will not suffice to address this question. The challenge lies in whether there is enough variation in admission criteria across centers for the data to yield useful insights. If the amount of "disagreement" is sufficient, analysis will focus on heterogeneity of effects, namely,
of studying the way admission changes prospects of a favourable clinical improvement and, potentially, aspects of causative interaction
between different decisions along the clinical journey, in the light of a global causal graph
representation of the therapeutic process.

\newpar This draft is the first version of a paper waiting for brain scan data to be made available.

\end{document}